\documentclass[twocolumn,showpacs,preprintnumbers,amsmath,amssymb]{revtex4}

\usepackage{graphicx}
\usepackage{dcolumn}
\usepackage{bm}

\begin{document}

\preprint{APS/123-QED}

\title{Critical Dynamics of Three-Dimensional Spin Systems
with Long-Range Interactions}

\author{S.V. Belim}
 \email{belim@univer.omsk.su}
\affiliation{%
Omsk State University, 55-a, pr. Mira, Omsk, Russia, 644077
\textbackslash\textbackslash
}%

\date{\today}

\begin{abstract}
A field-theoretic description of critical behavior of Ising
systems with long-range interactions is obtained by using the
Pade–Borel summation technique in the two-loop approximation
directly in the three-dimensional space. It is shown that
long-range interactions affect the relaxation time of the system.
\end{abstract}

\maketitle

It was shown in [1] that effects due to long-range interaction are
essential for the critical behavior of Ising systems. The
renormalization-group approach to spin systems with long-range
interactions developed in [2] directly in the three-dimensional
space made it possible to calculate the static critical exponents
in the two-loop approximation. However, analogous calculations of
critical dynamics have never been performed for these systems.

In this paper, a field-theoretic description of critical behavior
of homogeneous systems with long-range interactions is developed
directly for $D = 3$ in the two-loop approximation. The model
under analysis is the classical spin system with exchange integral
depending on the distance between spins. The corresponding
Hamiltonian is
\begin{equation}
H=\frac{1}{2}\sum_{ij}J(|r_i-r_j|)S_iS_j,
\end{equation}
where $S_i$ is a spin variable and $J(|r_i–r_j|)$ is the exchange
integral. This model is thermodynamically equivalent to the
$O(n)$-symmetric Ginzburg–Landau–Wilson model defined by the
effective Hamiltonian
\begin{equation}\label{gam1}
   H=\int d^Dq\Big\{\frac{1}{2}(\tau_0+q^\sigma)\varphi^2+u_0\varphi^4\Big\},
\end{equation}
where $\varphi$ is a fluctuating order parameter, $D$ is the space
dimension, $\tau_0\sim |T-T_c|$ ( $T_c$ is the critical temperature),
and $u_0$ is a positive constant. Critical behavior strongly depends
on the parameter a characterizing the interaction as a function of
distance. It was shown in [3] that long-range interaction is
essential when $0<\sigma<2$, whereas systems with $\sigma = 2$ exhibit
critical behavior characteristic of short-range interactions. For
this reason, the analysis that follows is restricted to the case
of $0<\sigma<2$.

Relaxational dynamics of spin systems near the critical
temperature can be described by a Langevin-type equation for the
order parameter:
\begin{equation} \label{Langevin}
\frac{\partial{\bf\varphi}}{\partial t}=-\lambda_0\frac{\delta
H}{\delta{\bf\varphi}}+{\bf\eta}+\lambda_0{\bf h},
\end{equation}
where $\lambda_0$ is a kinetic coefficient, ${\bf\eta}(x,t)$ is a gaussian
random force (representing the effect of a heat reservoir) defined
by the probability distribution
\begin{equation}
P_\eta=A_\eta\exp\left[-(4\lambda_0)^{-1}\int
d^{d}x\,dt\,{\bf\eta}^2(x,t)\right]
\end{equation}
with a normalization factor $A_\eta$, and ${\bf h}(t)$ is an external field
thermodynamically conjugate to the order parameter. The temporal
correlation function $G(x,t)$ of the order-parameter field can
be found by solving Eq. (3) with $H[{\bf\varphi}]$ given by (2)
for ${\bf\varphi}[{\bf\eta},{\bf h}]$ averaging the result over
$P_\eta$, and retaining the component
linear in ${\bf h}(0)$
\begin{equation}
G(x,t)=\frac{\delta}{\delta{\bf
h}(0)}[\langle{\bf\varphi}(x,t)\rangle]|_{h=0},
\end{equation}
where
\begin{equation}
[\langle{\bf\varphi}(x,t)\rangle]=B^{-1}\int
D\{{\bf\eta}\}{\bf\varphi}(x,t)P_\eta,
\end{equation}
\begin{equation}
B=\int D\{{\bf\eta}\}P_\eta.
\end{equation}

When applying the standard renormalization-group procedure to this
dynamical model, one has to deal with substantial difficulties.
However, it was shown in [4] that the model of critical dynamics
in homogeneous systems without long-range interaction based on a
Langevin-type equation is equivalent to that described by the
standard Lagrangian [5]
\begin{equation}
L=\int d^{d}x\,dt\,\left\{\lambda_0^{-1}{\bf\varphi}^2+
i{\bf\varphi}^*\left(\lambda_0^{-1}\frac{\partial{\bf\varphi}}{\partial
t}+ \frac{\delta H}{\delta{\bf\varphi}}\right)\right\},
\end{equation}
where $\varphi^*$ denotes an auxiliary field. The corresponding
correlation function $G(x,t)$ of the order parameter is defined
for a homogeneous system as
\begin{eqnarray}
&&G(x,t)=\langle{\bf\varphi}(0,0){\bf\varphi}(x,t)\rangle\\
&&=\Omega^{-1}\int D\{{\bf\varphi}\}
D\{{\bf\varphi}^*\}{\bf\varphi}(0,0){\bf\varphi}(x,t)\exp(-L[{\bf\varphi},{\bf\varphi}^*]),\nonumber
\end{eqnarray}
where
\begin{equation}
\Omega=\int
D\{{\bf\varphi}\}D\{{\bf\varphi}^*\}\exp(-L[{\bf\varphi},{\bf\varphi}^*]).
\end{equation}
Instead of dealing with the correlation function, it is reasonable
to invoke the Feynman diagram technique and represent the
corresponding vertex in the two-loop approximation as
\begin{eqnarray}
&&{\Gamma}^{(2)}(k,\omega;\tau_{0},u_{0},{\lambda}_{0})=
\tau_{0}+k^{\sigma}-\frac{i\omega}{{\lambda}_{0}}-96{u_0}^2D_0,\nonumber\\
&&D_0=\frac{3}{4}\int \frac{d^Dqd^Dp}
{(1+|\vec{q}|^\sigma)(1+|\vec{p}|^\sigma)}\\
&&\cdot \frac1{(3+|\vec{q}|^\sigma+|\vec{p}|^\sigma+
|\vec{p}+\vec{q}|^\sigma-i\omega/\lambda)}\nonumber
\end{eqnarray}
The next step in the field-theoretic approach is the calculation
of the scaling functions $\beta$, $\gamma_\tau$,
$\gamma_\varphi$ and $\gamma_\lambda$ in the
renormalization-group differential equation for vertexes
\begin{eqnarray}
&&\left[\mu\frac{\partial}{\partial\mu}+\beta\frac{\partial}{\partial
u} -\gamma_{\tau}\tau\frac{\partial}{\partial\tau}
+\gamma_{\lambda}\lambda\frac{\partial}{\partial\lambda}-
 \frac{m}{2}\gamma_\varphi\right]\nonumber\\
&&\times\Gamma^{(m)}(k,\omega;\tau,u,\lambda,\mu)=0,
\end{eqnarray}
where the scaling parameter $\mu$ is introduced to change to
dimensionless variables.

Further analysis requires the use of the function $\beta$ and the
dynamic scaling function $\gamma_{\lambda}$.

An expression for $\beta$ in the two-loop approximation were obtained in [2]:
\begin{eqnarray}
    \beta&=&-(4-D)\Big[1-36uJ_0+1728\Big(2J_1-J_0^2-\frac29G\Big)u^2)\Big],\nonumber\\
    J_1&=&\int \frac{d^Dq
    d^Dp}{(1+|\vec{q}|^\sigma)^2(1+|\vec{p}|^\sigma)
    (1+|q^2+p^2+2\vec{p}\vec{q}|^{\sigma/2})},\nonumber\\
    J_0&=&\int \frac{d^Dq}{(1+|\vec{q}|^\sigma)^2},\nonumber\\
    G&=&-\frac{\partial}{\partial |\vec{k}|^\sigma}\int \frac{d^Dq
    d^Dp}{(1+|q^2+k^2+2\vec{k}\vec{q}|^{\sigma/2})(1+|\vec{p}|^\sigma)}\nonumber\\
    &&\cdot\frac1{(1+|q^2+p^2+2\vec{p}\vec{q}|^{\sigma/2})}\nonumber
\end{eqnarray}
The function $\gamma_{\lambda}$ calculated in the two-loop approximation is
\begin{eqnarray}
    \gamma_\lambda=(2\sigma-D)2(D'-G)u^2,\\
    D'=\frac{\partial{D_0}}{\partial (-i\omega/\lambda)}|_{k=0,\omega =0}.\nonumber
\end{eqnarray}
Defining the effective interaction vertex $v=u\cdot J_0$
one obtains the following expressions for $\beta$ and $\gamma_{\lambda}$:
\begin{eqnarray}\label{beta}
    \beta&=&-(2\sigma-D)\Big[1-36v+1728\Big(2\widetilde{J_1}-1-\frac29\widetilde{G}\Big)v^2)\Big],\nonumber\\
    \gamma_\lambda&=&(2\sigma-D)96(\widetilde{D}-\widetilde{G})v^2,\\
    \widetilde{J_1}&=&\frac{J_1}{J_0^2}\ \ \ \
    \widetilde{G}=\frac{G}{J_0^2}\ \ \ \
    \widetilde{D}=\frac{D'}{J_0^2}.\nonumber
\end{eqnarray}
This redefinition is meaningful for $\sigma\leq D/2$. In this case,
$J_0$, $J_1$, $G$ and $D'$ are divergent functions. Introducing the cutoff
parameter $\Lambda$, we obtain finite expressions for the ratios
\begin{eqnarray}
&&\frac{J_1}{J_0^2}=\Big[\int_0^\Lambda\int_0^\Lambda d^Dq d^Dp/
((1+|\vec{q}|^\sigma)^2(1+|\vec{p}|^\sigma)\nonumber\\
&&(1+|q^2+p^2+2\vec{p}\vec{q}|^\sigma))\Big]\Big/
\Big[\int_0^\Lambda d^Dq/(1+|\vec{q}|^\sigma)^2\Big]^2,\nonumber\\
&&\frac{G}{J_0^2}=\Big[-\frac{\partial}{(\partial |\vec{k}|^\sigma)}
\int_0^\Lambda\int_0^\Lambda d^Dqd^Dp
/((1+|\vec{p}|^\sigma)\nonumber\\
&&(1+|q^2+k^2+2\vec{k}\vec{q}|^\sigma)
(1+|q^2+p^2+2\vec{p}\vec{q}|^\sigma))\Big]\nonumber\\
&&\Big/\Big[\int_0^\Lambda d^Dq/(1+|\vec{q}|^\sigma)^2\Big]^2,\\
&&\frac{D'}{J_0^2}=\frac34\Big[\int d^Dqd^Dp/
((1+|\vec{q}|^\sigma)(1+|\vec{p}|^\sigma)\nonumber\\
&&(3+|\vec{q}|^\sigma+|\vec{p}|^\sigma+
|\vec{p}+\vec{q}|^\sigma)^2)\Big]\Big/
\Big[\int_0^\Lambda d^Dq/(1+|\vec{q}|^\sigma)^2\Big]^2.\nonumber
\end{eqnarray}
as $\Lambda\rightarrow\infty$.

The integrals are performed numerically. For $\sigma\leq D/2$, a sequence
of $J_1/J_0^2$ and $G/J_0^2$ corresponding to various values of $\Lambda$
is calculated and extrapolated to infinity.

Critical behavior is completely determined by the stable fixed
points of the renormalization group (RG) transformation. These
points can be found from the condition
\begin{equation}\label{nep}
    \beta(v^*)=0.
\end{equation}
The effective interaction vertexes were evaluated at the stable
fixed points of the RG transformation in [2].

The dynamic critical exponent $z$ characterizing critical
slowing-down of relaxation is determined by substituting the
effective charges at a fixed point into the scaling function ãë:
\begin{equation}
z=2+\gamma_{\lambda}.
\end{equation}
The table shows the stable fixed points of the RG transformation
and the values of the dynamic critical exponent for $1.5\leq \sigma \leq 1.9$.
When $0<\sigma<1.5$, the only fixed point is the unstable gaussian
one, $v^* = 0$.

A comparison of the present results with the value $z = 2.017$ of
the dynamic critical exponent for three-dimensional systems with
short-range interactions obtained in [6] demonstrates the
essential role played by long-range interactions in critical
dynamics of spin systems. In particular, the system's relaxation
time increases according to the scaling $t\sim |T-T_c|^{\nu z}$, where $\nu$
is the critical exponent characterizing the increase in the
correlation radius near a critical point. In three-dimensional
systems with long-range interaction, both critical dynamics and
static behavior become increasingly gaussian as the long-range
interaction parameter $\sigma$ decreases. When $a\leq1.8$, critical
behavior is virtually gaussian.

The work is supported by Russian Foundation for Basic Research N 04-02-16002.

\begin{table*}
\hspace{120mm} {} \vspace{5mm}
Table 1.
\begin{center}
\begin{tabular}{ccc} \hline
 $a$   & $v^{*}$  &$z$ \\
\hline
 1.5 & 0.015151 &  2.000072  \\
 1.6 & 0.015974 &  2.000180  \\
 1.7 & 0.020485 &  2.000777  \\
 1.8 & 0.023230 &  2.001529\\
 1.9 & 0.042067 &  2.006628\\
 \hline
\end{tabular} \end{center} \end{table*}

\newpage
\def\baselinestretch{1.0}

\end{document}